\documentstyle[aps,pre,epsf]{revtex}

\begin{document}
\draft
\title{The Nonlinear Debye-Onsager Relaxation Effect in Weakly Ionized Plasmas}
\author{J. Ortner} 
\address{ Humboldt Universit\"at zu Berlin, Institut f\"ur Physik\\ Invalidenstr.110, D-10115 Berlin}
\date{Received 03 June 1997}

\maketitle

\begin{abstract}
A weakly ionized plasma under the influence of a strong electric field is considered. Supposing a local Maxwellian distribution for the electron momenta the plasma is described by hydrodynamic equations for the pair distribution functions. These equations are solved and the relaxation field is calculated for an arbitrary field strength. It is found that the relaxation effect becomes lower with increasing strength of the electrical field.
\end{abstract}

\section{Introduction}
Transport in high electric field has become a topic of current interest in 
various fields of physics. The shrinking sizes of modern devices with the applied 
voltages remaining the same produces very strong fields. A further motivation 
for the investigation of high field transport effects is the application of
strong laser fields to matter. Nonlinear transport effects due to strong short 
laserpulses can be studied in semiconductors \cite{1} and in plasma physics 
\cite{2}.

In this paper we consider a weakly ionized plasma in strong electric field. A plasma is called weakly  
ionized if the collision frequency of ions with neutrals is much greater than 
the ion-ion collision frequency. Therefore, in a weakly ionized plasma the 
relaxation time for the momenta of the charged particles is much less then 
the relaxation time for the screening cloud surrounding a charge. 

The influence of field effects to the plasma conductivity is twofold. First, 
the external field modifies the distribution function of electrons and induces
a current. Second, the field deforms the screening cloud. As a result an 
internal field is induced, which is directed opposite to the external field 
and diminishes the total field acting on the charges (the Debye-Onsager 
relaxation effect).

A kinetic equation for the determination of the electron distribution function 
in weakly ionized plasmas was derived by Dawydow \cite{3}. It was shown that 
the influence of the field on the electron distribution is weak as far as the 
electron energy gain by the field during the mean free path is much less then 
the energy transferred by one collision of the electron with a neutral, i.e., if

\begin{equation} \label{1}
\gamma = (eEl/T) \sqrt{(M/m)} \ll 1 \quad , 
\end{equation}
where e is the elementary charge, E the electric field strength, {\it l} the length of mean free path, T is the temperature of the plasma, m and M are the masses of the electrons and the neutrals, respectively.

The Debye-Onsager relaxation effect was first derived within the theory of 
electrolytes \cite{{4}}. It has been shown that in a dilute electrolyte 
solution with two sorts of ions the total field acting on a charged paricle is reduced by an amount: 
\begin{equation} \label{2}
\delta E = - E \frac{1}{3(1+\sqrt{q})} \frac{e_1e_2 q  k_D}{T \varepsilon} \quad ,
\end{equation}
where T and $\varepsilon$ are the temperature and the dielectric constant of the electrolyte solution, $k_D = \sqrt{4 \pi \sum_{a=1}^2 e_a^2 n_a /T}$ is the inverse Debye screening length, $q=(e_1b_1-e_2b_2)/(e_1-e_2)(b_1+b_2)$, $e_a$,  $n_a$ and $b_a$ being the charges, the number density and the mobility of the ions of type a.
Recently the Debye-Onsager relaxation effect has become a topic of renewed interest. First it was shown that the relaxation effect is essential for a proper virial expansion of the electrical conductivity of a fully ionized plasma \cite{Roepke}. Second the Debye-Onsager relaxation effect in fully ionized plasmas beyond linear response was studied within the memory approach \cite{Morawetz}.
The aim of this paper is to investigate the nonlinear Debye-Onsager relaxation effect in weakly ionized plasmas.
\section{Hydrodynamic Approximation}
Consider a weakly ionized plasma consisting of $N_0$ neutrals, $N_a$ particles of sort a with charges $e_a$, masses $M_a$, $N_b$ particles of sort b, etc. The total number of charged particles is $N=\sum_{a} N_a$, the neutrality condition reads $\sum_{a} e_a N_a =0$. We suppose that $N_0 \gg N$. Consider now the plasma under the influence of a static homogeneous electrical field $\vec{E}$. In what follows we suppose that the electric field strength is restricted by 
condition Eq. (\ref{1}) and the electron distribution function can be approximated by the Maxwell distribution. Besides Eq. (\ref{1}) no other restrictions will be imposed on the electrical field strength. As far as Eq.(\ref{1}) is satisfied we will regard the relation between the dipole energy of the deformed screening cloud to the temperature of the plasma as an arbitrary one. Of course we also neglect effects of impact ionization in an external field.

The system of charged particles is then described by the N-particle distribution function $F_N (x_1, \ldots,x_N, v_1, \ldots ,v_N)$, satisfying the generalized Liouville equation \cite{yklim}
\begin{eqnarray} \label{3}
\frac{\partial{F_N}}{\partial t} + \sum_{i=1}^{N} \vec{v_i} \frac{\partial{F_N}}{\partial{\vec{x_i}}} + \sum_{i=1}^{N} e_i \vec{E} \frac{\partial{F_N}}{\partial{\vec{v_i}}} &+& \sum_{i,j=1}^{N} \frac{1}{m_i} \frac{\partial{\Phi_{ij}(\vec{x_i} - \vec{x_j})}}{\partial{\vec{x_i}}} \frac{\partial{F_N}}{\partial{\vec{v_i}}}
\nonumber\\ 
    &=&  St F_N = \sum_{i=1}^N {St}_i(v_i) F_N \quad  . 
\end{eqnarray}

The right-hand side  of Eq. (\ref{3}) describes the collisions of the charged particles with the neutrals. In Eq. (\ref{3}) we suppose that the collisions of two different particles are independent. Introducing truncated distribution functions one obtains from Eq. (\ref{3}) the Bogoliubov-Born-Green-Kirkwood-Yvon (BBGKY) hierarchy of kinetic equations \cite{9}. If the charged component of the plasma forms a weakly coupled plasma, i.e., if  
\begin{displaymath}
({e^2}/{T}) {\left( {4 \pi N}/{3V} \right)}^{1/3} \ll 1 \quad , \nonumber
\end{displaymath}
the BBGKY hierarchy can be truncated by supposing that all distribution functions of order higher than 2 can be expressed by the one- and two-particle distribution functions. Then one arrives at a system of equations for the one- and two-particles distribution functions \cite{9,yklim} 

This system can be further simplified by supposing a local Maxwellian distribution for the velocities in the one and two-particle distribution functions. In the quasistationary case one arrives thus at the system of equations \cite{yklim}
\begin{equation} \label{5}
\Delta_a \phi_b(\vec{r_a},\vec{r_b}) = - 4 \pi  {\left[\sum_{c}e_c N_c h_{cb}(\vec{r_a},\vec{r_b}) + e_b \delta (\vec{r_a}-\vec{r_b}) \right]} \quad ,
\end{equation}
\begin{eqnarray} \label{6}
T(b_a+b_b) \Delta h_{ab}(\vec{r_a},\vec{r_b}) + e_a b_a \Delta_a \phi_b(\vec{r_a},\vec{r_b})+ e_b b_b \Delta_a \phi_a(\vec{r_b},\vec{r_a}) &=& \nonumber \\
{\left( e_ab_a - e_b b_b \right)} \vec{E} \cdot \vec{\nabla} h_{ab}(\vec{r_a},\vec{r_b}) \quad &.&
\end{eqnarray}
In Eq. (\ref{6}) $b_a$ is the mobility of a particle of sort a, $\phi_b(\vec{r_a},\vec{r_b})$ is the potential created at point $ \vec{r_a}$ by a particle of sort b situated at point  $ \vec{r_b}$ and $h_{ab}(\vec{r_a},\vec{r_b})$ is the two-particle distribution function. 
The system of differential equations Eqs. (\ref{5}) and (\ref{6}) can be solved by a Fourier transformation. For the sake of simplicity we consider now a weakly ionized plasma consisting of neutrals, electrons and one sort of ions with charges $Ze$. Since the mobility of the electrons is much greater than that of the ions we neglect the ion mobility. Than a straightforward calculation leads to the following expression for the Fourier transforms of the distribution functions and the potentials:
\begin{eqnarray} \label{7}
\tilde{h}_{ie}(\vec{k})&=& \tilde{h}_{ei}(-\vec{k}) = \frac{4 \pi Z e^2}{T(k^2+k_D^2)} \frac{1-\frac{e}{T} \frac{i \vec{k} \cdot \vec{E}}{k^2+k_{De}^2}}{1-x} \qquad , 
\nonumber\\ 
\tilde{h}_{ii}(\vec{k}) &=& - \frac{4 \pi Z^2 e^2}{T(k^2+k_D^2)} \frac{1- x \frac{k^2+k_D^2}{k^2+Z k_{De}^2}}{1-x} \quad , \qquad
\tilde{h}_{ee}(\vec{k}) = - \frac{4 \pi  e^2}{T(k^2+k_D^2)} \frac{1- x \frac{k^2+k_D^2}{k^2+ k_{De}^2}}{1-x} \quad ,  
\nonumber\\
\tilde{\varphi_e}(\vec{k})&=& \frac{4 \pi  e}{k^2+k_D^2} \frac{1}{1-x} {\left[-1-\frac{Ze}{T} \frac{i \vec{k} \cdot \vec{E}}{k^2 + k_{De}^2} \frac{k_{De}^2}{k^2} + x \frac{k^2+k_D^2}{k^2+ k_{De}^2}    \right]} \quad ,
\nonumber\\
\tilde{\varphi_i}(\vec{k}) &=& \frac{4 \pi Z  e}{k^2+k_D^2} \frac{1}{1-x} {\left[1-\frac{e}{T} \frac{i \vec{k} \cdot \vec{E}}{k^2 + k_{De}^2} \frac{k_{De}^2}{k^2} - x \frac{k^2+k_D^2}{k^2+ Z k_{De}^2}    \right]} \quad ,   
\end{eqnarray}
where
\begin{displaymath}
x= { \left( \frac{e}{T} \right) }^2 \frac {(i \vec{k} \cdot \vec{E} )^2}{k^2(k^2+k_D^2)} \frac {k^2+Z k_{De}^2}{k^2+k_{De}^2}  \quad , \qquad k_{De}^2 = \frac{4
\pi n_e e^2}{T} \quad , \qquad k_D^2 = (Z+1)k_{De}^2 
\end{displaymath}
$\varphi_e(\vec{r} , \vec{r_e})$ is the additional potential at point $\vec{r}$ if an electron is situated at point $\vec{r_e}$. The field strength of this potential equals
\begin{displaymath}
\vec{E}_e(\vec{r} - \vec{r_e}) = - \vec{\nabla} \varphi_e (\vec{r} , \vec{r_e})= - \vec{\nabla} \varphi_e (\vec{r} - \vec{r_e}) \quad .
\end{displaymath}

Putting now $\vec{r}=\vec{r_e}$ we obtain the potential acting on the electron itself and thus changing its mobility. For the Fourier transform we have $\vec{E_e}(\vec{k}) = - i \vec{k}\varphi_e(\vec{k})$. Therefore
\begin{eqnarray} \label{8}
\delta \vec{E} &=& \vec{E_e}(\vec{r}=0) = - \int i \vec{k}\varphi_e(\vec{k})\, \frac {d^3 k}{(2 \pi)^3}
\nonumber\\
&=& - \frac {Z e^2 k_{De}}{T} \vec{E} \int_0^1 dy \, y^2 {\left( 2 + Z+ \alpha^2 y^2 + 2 \sqrt{Z+1+Z\alpha^2 y^2} \right)}^{-1/2} \quad ,
\end{eqnarray} 
with $\alpha=eE/Tk_{De}$ characterizing the strength of the electrical field. For the case of onefold charged ions ($Z=1$) the integral in Eq. (\ref{8}) can be expressed by elementary functions, and one obtains
\begin{eqnarray} \label{9}
\delta \vec{E}= - \frac { e^2 k_{De}}{3(1+\sqrt{2}) \, T} &\vec{E}&\, F(\alpha) \quad ,\nonumber\\
 F(\alpha) &=& \frac{3(1+\sqrt{2})}{\alpha^2} {\left[ \frac{1}{2} \sqrt{\alpha^2 + 2} - 1 + \frac{1}{\alpha}\arctan(\alpha) - \frac{1}{\alpha}\arctan(\frac{\alpha}{\sqrt{\alpha^2 + 2}})
\right]} \quad .
\end{eqnarray} 
This formula is the main result of the present paper. It will be discussed in the next section.

\section{Discussions}
The nonlinear relaxation effect can be described via the function $F(\alpha)$ in Eq. (\ref{9}). For the case of a weak external electric field $\alpha \ll 1$ the relaxation field differs from the corresponding field in the linear regime by a small amount, and is given by the expression
\begin{equation} \label{10}
F(\alpha) = 1 - \frac{3 \alpha^2}{20 ( \sqrt{2} + 1)} \quad .
\end{equation}

In the opposite case of a very strong electric field $\alpha \gg 1$ the relaxation effect is described by the asymptotic behavior of the function $F(\alpha)$,
\begin{equation} \label{11}
 F(\alpha) \sim  \frac{3  ( \sqrt{2} + 1)}{2 \alpha }\quad .
\end{equation}

One can compare the relaxation effect calculated in the present approach with the corresponding result from a memory approach \cite{Morawetz}. In the latter approach a kinetic equation for the one-particle distribution function is employed which differs from the ordinary Boltzmann equation in two points. (1) A collisional broadening as a result of the finite collision duration. (2) The intracollisional field effect, which gives additional retardation effects in the momentum of the one-particle distribution function due to the field. Then the relaxation field is found to be
\begin{equation} \label{12}
\delta \vec{E}= - \frac { e^2 k_{De}}{3 \sqrt{2} T} \vec{E}\, F_M(\alpha) \quad , \qquad
 F_M(\alpha) = \frac{6}{\alpha^2} {\left[- \frac{3}{2} + \frac{\alpha}{2} + \frac{2}{\alpha}\ln{(1+\alpha)} - \frac{1}{2(\alpha + 1)}
\right]} \quad .
\end{equation}
(Unfortunately in \cite{Morawetz} is given an incorrect result, in Eq (\ref{12}) we have corrected the result.) Note that the memory approach leads to another expression for the linear relaxation effect than the hydrodynamic approach.

  In Fig. 1 we have plotted the nonlinear relaxation effect via the functions $F(\alpha)$ from Eq. (\ref{9}) and $F_M(\alpha)$ from Eq. (\ref{12}). One regognizes that within the both approaches the relaxation effect lowers with increasing strength of the external electric field and that for very strong external fields the relaxation field goes asymptotically to zero. Within the memory approach a faster decay is predicted.

The vanishing of the relaxation effect can be understood as follows. In a strong external electric field the electron-ion plasma decouples and forms two almost independent subsystems, an electron and an ion subsystem moving in opposite directions. As a result the electron-ion correlations [described by $\tilde{h}_{ie}(\vec{k})$ in Eq. (\ref{7})] vanish, there is no deformed screening cloud anymore and the relaxation effect is absent.

\begin {figure} [h]
  \begin{center}
    \leavevmode
    \epsfxsize=70mm
    \epsffile{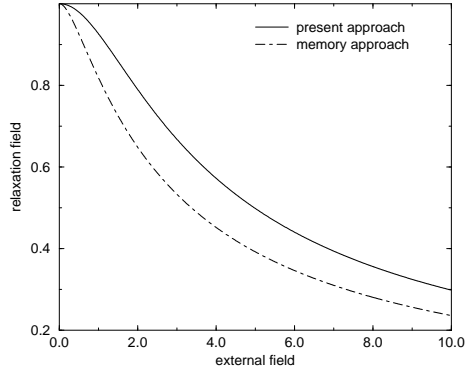}
   \end{center}
\caption{The scaled nonlinear Debye-Onsager relaxation field  $F(\alpha)$ (Eq(\ref{9})) and $F_M(\alpha)$ (Eq(\ref{12})) versus the field parameter $\alpha = e^2 E/ k_{De} T$.} 
\end{figure}

\section{Conclusions}
We have examined the nonlinear Debye-Onsager relaxation effect in weakly coupled plasmas. The hydrodynamic approximation was used in order to derive equations for the pair correlation functions and the relaxation field. It was shown that with increasing external field the relaxation field becomes lower and goes asymptotically to zero for the case of very strong electric fields.
    
We should remember here that the asymptotic region $\alpha \gg 1$ is rather a hypothetical one since at very strong external fields the condition Eq. (\ref{1}) breaks down and the hydrodynamic approach used in our calculations becomes inapplicable. On the other hand the hydrodynamic equations for the pair correlation functions Eqs. (\ref{7}) are also valid for electrolytes, and they should be valid, too, if a strong external electric field is applied. However, the present approach does not take into account electrophoretic effects which become important for electrolytes \cite{Falkenhagen}.

\section*{Acknowledgments}
The author acknowledges stimulating discussions with K. Morawetz. I also thank Dr. Morawetz for making me familiar with Ref. \cite{Morawetz} before publication. This work was supported by the DFG (Germany).

\end{document}